\documentclass[twocolumn,prl,aps]{revtex4}

\usepackage{amssymb}
\usepackage{amsfonts}
\usepackage{amsmath}
\usepackage{bm}
\usepackage{graphicx}        
\usepackage[usenames,dvipsnames]{xcolor}
\usepackage{changes}
\usepackage{siunitx}
\usepackage{gensymb}
\usepackage{mathtools}
\usepackage[latin1]{inputenc}
\usepackage{CJK}
\usepackage[section,numberedsection=autolabel]{glossaries}
\usepackage[linkcolor=violet,citecolor=blue,colorlinks=true,urlcolor=purple]{hyperref}

\def\cep{\varphi_\textrm{CEP}}
\def\asymm{\mathcal{A}_z}

\makeglossaries
\newacronym{cep}{CEP}{carrier-envelope phase}
\newacronym{req}{$R_{eq}$}{equilibrium internuclear distance}
\newacronym{lite}{LITE}{laser-induced transfer of electron}
\newacronym{CT}{CT}{charge-transfer}
\newacronym{vdw}{vdW}{van der Waals}

\newcommand{\Sa}[1]{\textsf{#1}}

\renewcommand{\vec}{\textbf}
\newcommand{\SM}{{Suppl. Mat. \cite{SMWang2020}}}

\begin{document}

\begin{CJK*}{UTF8}{gbsn}

\title{Laser-induced Electron-Transfer in the Dissociative Multiple Ionization of Argon Dimers}

\author{YanLan\,Wang, XuanYang\,Lai, ShaoGang\,Yu, RenPing\,Sun, XiaoJun\,Liu}\email[]{xjliu@wipm.ac.cn}
\affiliation{State Key Laboratory of Magnetic Resonance and Atomic and Molecular Physics, Innovation Academy for Precision Measurement Science and Technology, Chinese Academy of Sciences, Wuhan 430071, China}

\author{Martin\,Dorner-Kirchner$^a$, Sonia\,Erattupuzha$^a$, Seyedreza\,Larimian$^a$, Markus\,Koch$^b$, V\'{a}clav\,Hanus$^a$, Sarayoo\,Kangaparambil$^a$, Gerhard\,Paulus$^c$, Andrius\,Baltu\v{s}ka$^a$, Xinhua\,Xie (谢新华)$^{a,d}$, Markus\,Kitzler-Zeiler$^a$}\email[]{markus.kitzler-zeiler@tuwien.ac.at}
\affiliation{$^a$Photonics Institute, Technische Universit\"at Wien, A-1040 Vienna, Austria,\\
$^b$Institute of Experimental Physics, Graz University of Technology, A-8010 Graz, Austria,\\
$^c$Institute of Optics and Quantum Electronics, Friedrich Schiller University Jena, D-07743 Jena, Germany,\\
$^d$SwissFEL, Paul Scherrer Institute, 5232 Villigen PSI, Switzerland
}

\begin{abstract}
We report on an experimental and theoretical study of the ionization-fragmentation dynamics of argon dimers in intense few-cycle laser pulses with a tagged carrier-envelope phase.
We find that a field-driven electron transfer process from one argon atom  across the system boundary to the other argon atom triggers sub-cycle electron-electron interaction dynamics in the neighboring atom.
This attosecond electron-transfer process between distant entities and its implications manifest themselves as a distinct phase-shift between the measured asymmetry of electron emission curves of the $\text{Ar}^{+}+\text{Ar}^{2+}$ and $\text{Ar}^{2+}+\text{Ar}^{2+}$ fragmentation channels.
Our work discloses a strong-field route to controlling the dynamics in  molecular compounds through the excitation of electronic dynamics on a distant molecule by driving inter-molecular electron-transfer processes.
\end{abstract}


 \maketitle

\end{CJK*}

Photoinduced molecular \gls{CT}
across system boundaries is a key step
in many important natural or technical processes 
such as
solar-driven energy production \cite{Yella2011, Lu2018}, photocatalysis   \cite{Morawski2014, Karkas2014}, or photosynthetic activity \cite{Barber2009, Rochaix2011}.
%
%
In these processes the relocation of charge, initiated by the absorption of a single
photon
by a molecule, is determined by the energetic and spatial structure of the system.
%
A fundamentally different mechanism for determining charge-localization
processes becomes available 
in strong laser fields. 
It was shown
that the \textit{intra-molecular} localization of electrons during the dissociation of isolated, small molecules can be determined by multi-photon processes driven by intense few-cycle laser pulses using their 
\gls{cep} as the control parameter \cite{Kling2006, Kling2013, Kremer2009, Fischer2010, Znakovskaya2012}. 

An intriguing yet unexplored question is then, whether strong-field-driven multi-photon processes can influence the localization of charge not only within one molecule but also \textit{across} system boundaries.
Widely used model-systems for investigating 
inter-system transfer reactions are small \gls{vdw} clusters and dimers.
\Gls{vdw} dimers are used to study photoinduced biological processes  \cite{Lippert2004_PCCP, Schultz2004, Horke2016}, photocatalytic reactions  \cite{Ehrmaier2017, Rabe2018}, and energy or charge transfer reactions induced by soft X-ray photons \cite{Cederbaum1997, Jahnke2004b, Sisourat2010, Mizuno2017} and electron impact \cite{Ren2016}.
\Gls{vdw} systems are also studied with strong laser fields, but in the case of dimers with a focus on the field-driven ionization and fragmentation dynamics \cite{Manschwetus2010, Ulrich2010, Wu2011, Ulrich2011, Wu2011b, Wu2012d, Hoshina2012, Wu2012b, VonVeltheim2013, Wu2014a, Amada2015, Ding2017, Cheng2017, Bogomolov2017}, or electronic 
 energy conversion processes in the case of larger clusters \cite{Ditmire1997, Kumarappan2001, Jungreuthmayer2004_PRL, Fennel2010, Schutte2015, Schutte2018}.
To the best of our knowledge, strong-field driven electron transfer-reactions across the system boundary from one entity to another have thus far not been investigated.



In this Letter, we show experimentally and by simulations, using the argon dimer, Ar$_{2}$, as an example, that electron transfer-reactions from one argon atom to the other can be driven by a strong laser field and, furthermore, that they are decisive for the
ionization and fragmentation behavior of the dimer.
%
Specifically, we demonstrate that an electron liberated at one of the two Ar atoms 
can be captured by the neighboring atom.
This process, which we refer to as the \gls{lite} process, determines the emission timing of the electrons via electron-electron interaction and thus,
depending on the \gls{cep}, influences the momenta of the emitted electrons.
As a result, the effect of \gls{lite} can be observed in our experiments and simulations
when comparing the asymmetry of electron emission as a function of \gls{cep} for the two ionization-fragmentation channels Ar(1,2) and Ar(2,2), where Ar($n,m$) denotes $\text{Ar}_2  \xrightarrow{\text{Laser}} \text{Ar}_{2}^{(n+m)+} \rightarrow \text{Ar}^{n+}+\text{Ar}^{m+}$.

In our experiments, argon dimers created by supersonic expansion of a few bars of argon gas were ionized by intense laser pulses, linearly polarized along $z$, with a full width at half maximum (FWHM) duration in intensity of 4.5\,fs
and a peak intensity, calibrated in \textit{in situ} \citep{Smeenk2011}, of \SI{5e14}{\watt\per\cm\square}, inside the ultra-high vacuum chamber of a reaction microscope \cite{Doerner2000}.
Details on the reaction microscope can be found in Refs.\,\cite{Xie2012_CE, Zhang2014a, Xie2017b}.
The laser center wavelength was $\lambda=750$\,nm. 
The duration of the pulses and their \gls{cep} were measured with a stereo electron spectrometer in phase-tagging mode \citep{Rathje2012}.
Upon laser ionization of the argon dimers, the two \gls{vdw}-bound argon atoms, separated by their \gls{req}
undergo fragmentation via Coulomb explosion.
We detected the two emerging argon ions, Ar$^{n+}$ and Ar$^{m+}$, in coincidence and from their time of flight and impact position on our detector calculated their three-dimensional momenta $\vec{p}_\text{Ar}^{n}$ and $\vec{p}_\text{Ar}^{m}$. By imposing momentum conservation conditions onto the ions detected in coincidence, the two-body fragmentation channels of interest, Ar(1,2) and Ar(2,2), as well as the channel Ar(1,1), were selected for further analysis.
Due to momentum conservation, the sum momentum of the $(n+m)$ emitted electrons,  $\vec{p}_e^{(n,m)} = \sum_{i=1}^{n+m} \vec{p}_{ei}$, with $\vec{p}_{ei}$ the momentum of the $i^\text{th}$ electron,
can be determined from the center of mass recoil momentum of the ions,
$\vec{p}_\text{R}^{(n,m)} = \vec{p}_\text{Ar}^{n}+ \vec{p}_\text{Ar}^{m}$,
using the relation
$\vec{p}_e^{(n,m)} = -\vec{p}_\text{R}^{(n,m)}$.
%

\begin{figure}[t]
	\centering\includegraphics[width=\columnwidth]{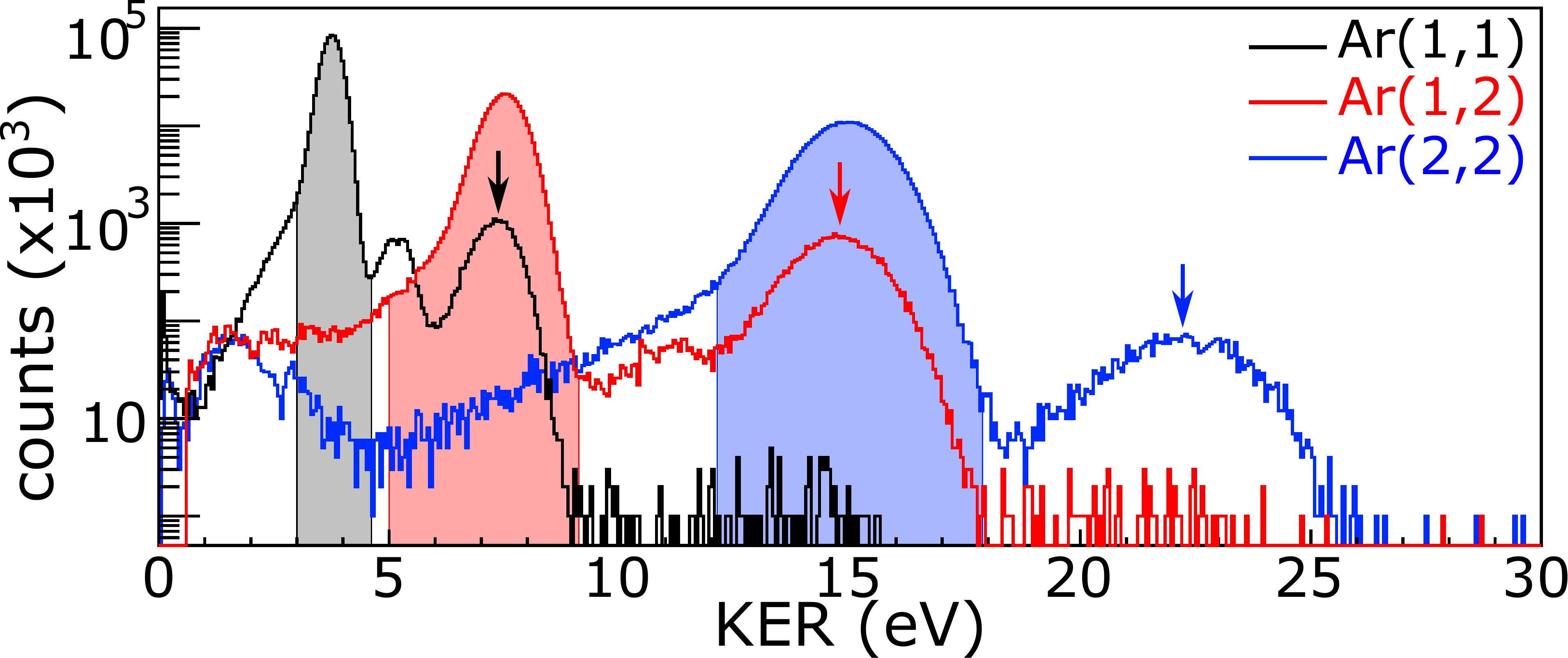}
	\caption{Kinetic energy release (KER) distributions of fragmentation channels Ar($n,m$) with $(n,m)=\{(1,1),(1,2),(2,2)\}$. 
Arrows mark peaks due to electron recapture, shaded areas highlight the peaks resulting from Coulomb explosion at \gls{req}.
		\label{fig1}}
\end{figure}

\begin{figure*}[t]
\centering\includegraphics[width=\textwidth]{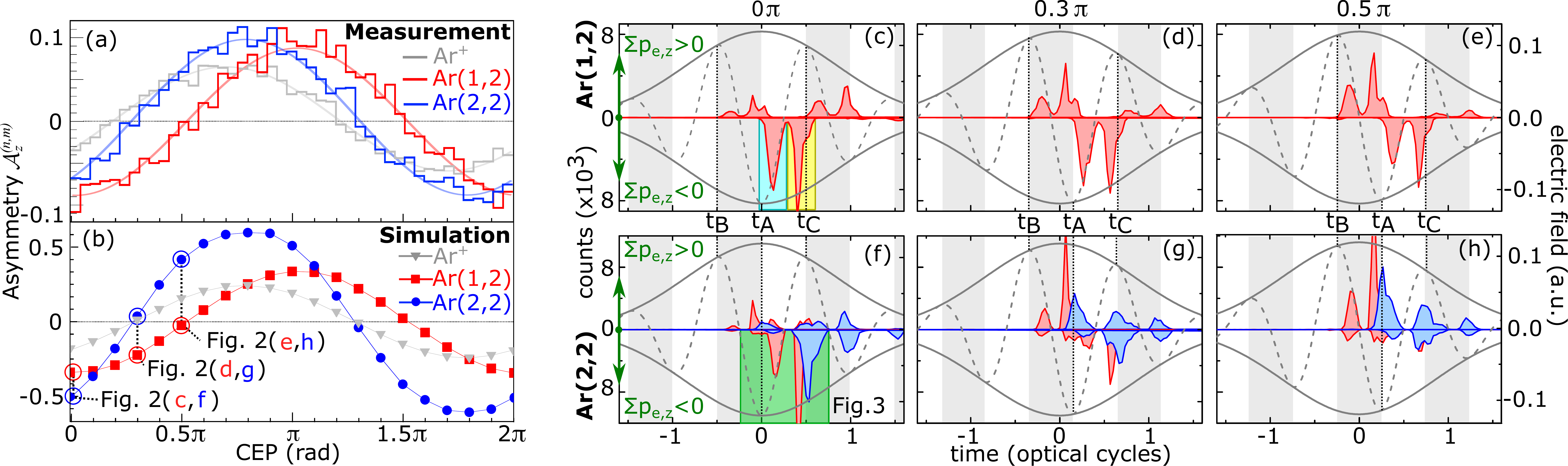}
\caption{(a) Measured asymmetry $\asymm^{(n,m)}$ of electron emission along $z$ for Ar(n,m), $(n,m)=\{(1,2),(2,2)\}$ over \gls{cep}.
(b) Same as (a) but simulated.
The curves of the Ar$^+$ monomer (gray) serve as a reference in (a) and (b).
(c-e) Simulated distributions of ionization times of the first electron, $t_{1st}$, for trajectories leading to Ar(1,2) for three values of the \gls{cep}.
(f-h) Same as (c-e) but for trajectory pairs leading to Ar(2,2) with the distributions of the ionization times of the second electron, $t_{2nd}$, shown in blue.
The laser electric field $E_{z}(t)$ (gray dashed) is also shown for reference.
The time-distributions are separated depending on whether the (sum) momentum of the electron (pair) reaches positive (upper halves) or negative momentum (lower halves).
\label{fig2}}
\end{figure*}

Fig.~\ref{fig1}(a) displays the measured distributions of the kinetic energy released (KER) during fragmentation, $\text{KER} = \left[(\vec{p}_\text{Ar}^{m})^2+(\vec{p}_\text{Ar}^{n})^2\right]/(2M)$ with $M$ the atomic mass of argon, for the Ar(1,1), Ar(1,2) and Ar(2,2) channels.
For each channel, at least two characteristic main peaks can be identified. The smaller peaks at higher KER-values (marked by arrows) were attributed to the process of frustrated tunnel ionization
\cite{Manschwetus2010, Ulrich2010, Wu2011}.
The dominant peaks at lower KER-values, highlighted by colored areas in Fig.~\ref{fig1}(a), originate from Coulomb explosions of the argon dimers at  \gls{req} and are the focus of this work.

%
To obtain insight into the 
multiple ionization dynamics
underlying the colored lower-KER peaks in Fig.~\ref{fig1}(a), 
we introduce an asymmetry-parameter $\asymm^{(n,m)} =(n_\text{up} - n_\text{dn})/(n_\text{up} + n_\text{dn})$, where $n_\text{up}$ ($n_\text{dn}$) denote for the channel Ar(n,m) the number of events with a positive (negative) electron sum momentum along $z$.
Alternatively to $\asymm^{(n,m)}$, one could also analyze the mean electron sum momentum $\bar{\vec{p}}_{e,z}^{(n,m)}$. But as we show in \SM{}, the two quantities feature an almost identical dependence on the CEP.
In the following we will use $\asymm^{(n,m)}$, as it has the advantage that it can be visually connected to electron yields discussed below. 
The measured dependence of $\asymm$ on \gls{cep} for the Ar(1,2) and Ar(2,2) channels is depicted  in Fig.~\ref{fig2}(a).
The key feature in Fig.~\ref{fig2}(a) is that the $\asymm$-curve for the Ar(2,2) channel exhibits a clear left phase shift of about
$0.23\pi$ to that of the Ar(1,2) channel.

To understand this experimentally observed \gls{cep}-shift between the two channels, we traced the correlated electrons and the motion of the nuclei in the combined laser and Coulomb fields by performing a 3D classical ensemble model calculation \cite{Xie2015d,Cheng2017},  described in \SM{}.
As the laser intensity is well above the over-the-barrier threshold \cite{Scrinzi1999} the two outermost electrons are rapidly stripped from each argon atom \cite{Xie2015d}. We therefore did not model these two initial ionization events and instead started from a dimer consisting of two singly charged argon ions (Ar$^{+}$- Ar$^{+}$), with one active electron situated around the position of each ion. 

The \gls{cep}-dependence of $\asymm$ predicted by the simulations for the Ar(1,2) and Ar(2,2) channels is shown in Fig.~\ref{fig2}(b).
%
The simulated curves agree very well with the measured ones, in particular the 
CEP left-shift 
of the Ar(2,2) channel 
is very well reproduced. 
The origin of this phase-shift can be extracted from the simulations by analyzing the distributions of ionization times $t_{1st}$ and $t_{2nd}>t_{1st}$ of the laser-driven electron trajectories that lead to the channels Ar(1,2) and Ar(2,2),
respectively.
The ionization time $t_{1st}$ marks the instant at which the single-particle energy of the first emitted electron becomes positive for the first time. Likewise, $t_{2nd}$ marks this instant for the second emitted electron in the Ar(2,2) channel.
The distributions of $t_{1st}$ and $t_{2nd}$ are plotted in Figs.~\ref{fig2}(c)-(h) for the Ar(1,2) and Ar(2,2) channels and three selected values of the \gls{cep}.
For convenience of the following discussion, the ionization time-distributions were separated depending on whether the (sum) momentum of the electron (pair) reaches positive (upper halves) or negative momentum (lower halves).


To explain the CEP left-shift between Ar(1,2) and Ar(2,2), we start with the CEP-dependence of $\asymm^{(1,2)}$.  
As shown  in Figs.~\ref{fig2}(c)-(e), the distributions of the ionization times ($t_{1st}$) in this channel feature two maxima per peak of the laser field. 
The reason underlying these two maxima will be explained below. 
For $\cep=0$, the two maxima corresponding to the field peak at \bm{$t_A$} are marked by cyan and yellow boxes [Fig.~\ref{fig2}(c)]. 
The maxima corresponding to the field peaks at \bm{$t_B$} and \bm{$t_C$} are much smaller for $\cep=0$. 
The  emission directions of electrons  set free during these maxima (up or down, indicated by positive or negative time-distributions) are largely determined by the laser vector potential according to the relation $\vec{p}_{e1}=-\vec{A}(t_i)=\int_{-\infty}^{t_i} \vec{E}(t') dt'$ \cite{Faisal1973, Reiss1980} with $t_i$ the ionization time. Positive values of $\vec{A}(t_i)$ are  indicated by gray shading in Figs.~\ref{fig2}(c)-(h). The small deviations from  $\vec{p}_{e1}=-\vec{A}(t_i)$ are due to the Coulomb forces of the argon ions.

For $\cep=0$, most of the trajectories are emitted with $\vec{p}_{e1}<0$. Therefore, $\asymm^{(1,2)}$ has a large negative value, cf. Figs.~\ref{fig2}(a,b).
For increasing CEP, the laser field-maximum at \bm{$t_B$} shifts closer to the pulse peak and becomes stronger. Accordingly, the positive valued double-peak structure corresponding to the field maximum at \bm{$t_B$} becomes gradually larger; at $\cep=0.5\pi$ the negative and positive double-peak structures are roughly equal in area. 
As a consequence, $\asymm^{(1,2)}$ varies from a large negative value at $\cep=0$ to roughly 0 at $\cep=0.5\pi$. Thus, the CEP-dependence of $\asymm^{(1,2)}$ in Figs.~\ref{fig2}(a,b) can to a good degree be explained straightforwardly using standard strong-field arguments based on the relation $\vec{p}_{e1}=-\vec{A}(t_i)$  and the sub-cycle dependence of the ionization rate on CEP.

To explain the CEP-dependence of $\asymm^{(2,2)}$ for Ar(2,2), also the distribution of $t_{2nd}$, blue-colored in Fig.~\ref{fig2}, must be considered. The distributions of $t_{1st}$ in the Ar(2,2) channel, although different in amplitude from those of channel Ar(1,2), are also dominated by two peaks per laser cycle. In contrast, the distribution of $t_{2nd}$ for $\cep=0$ in Fig.~\ref{fig2}(f) is dominated by only one peak. It is delayed by a laser-half-cycle to the strongest field maximum at \bm{$t_A$} and points into the negative direction. 
Again, the  reasons for the delay and the single peak-structure will be discussed below. 
Together with the $t_{1st}$ peaks that also point into the negative direction, this single $t_{2nd}$ peak leads to $\asymm^{(2,2)}<0$ for $\cep=0$, in agreement with Figs.~\ref{fig2}(a,b). 
As the CEP increases, the half-cycle-delayed negative $t_{2nd}$ peak due to the decreasing field-maximum at \bm{$t_A$} becomes weaker, and the positive $t_{2nd}$ peak due to the increasing field maximum at \bm{$t_B$} becomes stronger. 
Together with the $t_{1st}$ distributions that behave similarly as in the Ar(1,2) case, this causes that $\asymm^{(2,2)}$ moves towards positive values, reaches $\approx 0$ for $\cep=0.3\pi$ and a large positive value for $\cep=0.5\pi$ [see Figs.~\ref{fig2}(a,b)].

\begin{figure}[t]
\centering
\includegraphics[width=\columnwidth]{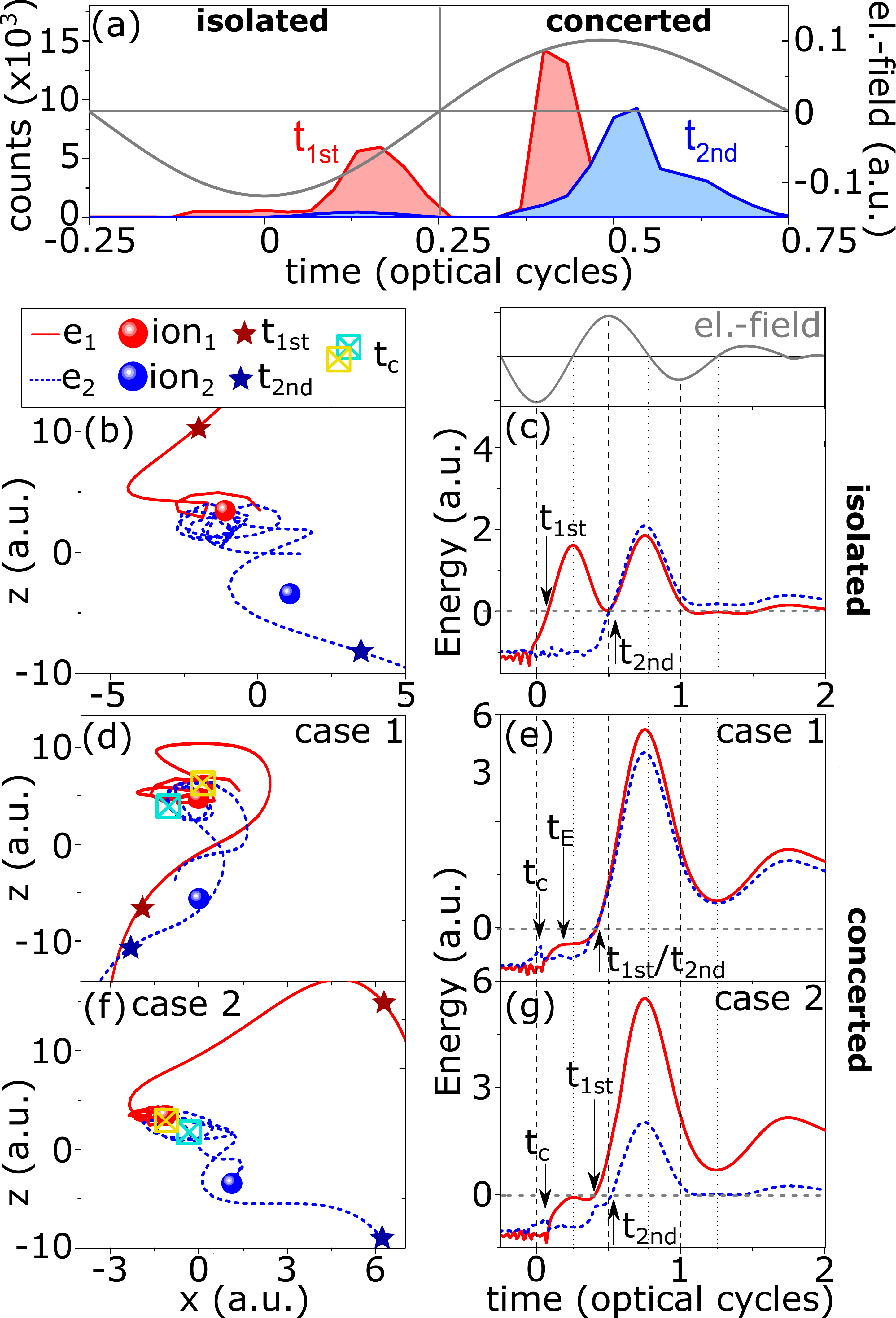}
\caption{Classical trajectory analysis for channel Ar(2,2). (a) Ionization time distributions of first (red), $t_{1st}$, and second (blue), $t_{2nd}$, electrons for electron pairs emitted within [-0.25T, 0.75T] and with negative sum momentum for $\cep=0$.
The left (b,d,f) and right columns (c,e,g) show typical electron trajectories in space and over time, respectively. The trajectories are classified in isolated ($t_{1st}\in[-0.25T, 0.25T]$, (b,c)) and concerted ($t_{1st}\in[0.25T, 0.75T]$, (d-g)). \Sa{t}$_{\Sa{C}}$, \Sa{t}$_{\Sa{E}}$ denote the times of collision and excitation. For better visibilty, the orbits in (b,d,f) are shown for $t>-0.1T$. \label{fig3}}
\end{figure}

We now turn to discussing the origin of the $t_{1st}$ double-peak and the half-cycle delayed single-peak structure of $t_{2nd}$. As we will see, this will also explain the CEP left-shift of $\asymm^{(2,2)}$ relative to $\asymm^{(1,2)}$. 
%
To this end, we traced the classical trajectories leading to the Ar(2,2) channel. 
For simplicity, but without loss of generality,
we select for this in-depth analysis the electron pairs emitted within $[-0.25T, 0.75T]$ and with negative sum momentum for the case of $\cep=0$ [indicated by a green box in Fig.~\ref{fig2}(f)].
The resulting time-distributions, displayed in Fig.~\ref{fig3}(a), show that the emissions can be classified into two types according to the relative emission time of the first and second electrons: One, where the two emissions happen \textit{isolated} of each other ($t_{1st} \in [-0.25T, 0.25T]$), and a second one, where the two emission steps happen in a \textit{concerted} manner within the same half-cycle ($t_{1st} \in [0.25T, 0.75T]$).

Typical trajectories for both the \textit{isolated} and \textit{concerted} case are displayed in Fig.~\ref{fig3}. 
We show in the \SM{} that they are representative for all emitted trajectories.
The trajectories for the isolated case of double ionization (DI) [Figs.~\ref{fig3}(b,c)] show that the first electron is immediately flying away from its own parent nucleus (red curve). The second electron, in contrast, is transferred to the other nucleus, where it is subsequently temporally captured by the Coulomb potential of the neighboring Ar$^{2+}$ ion. It becomes ionized only during the next laser-half-cycle around the peak of the field.
We refer to this electron transfer process across the system boundaries and the subsequent capture process that results in the ionization delay as \gls{lite}.

\begin{figure}[t]
\centering
\includegraphics[width=\columnwidth]{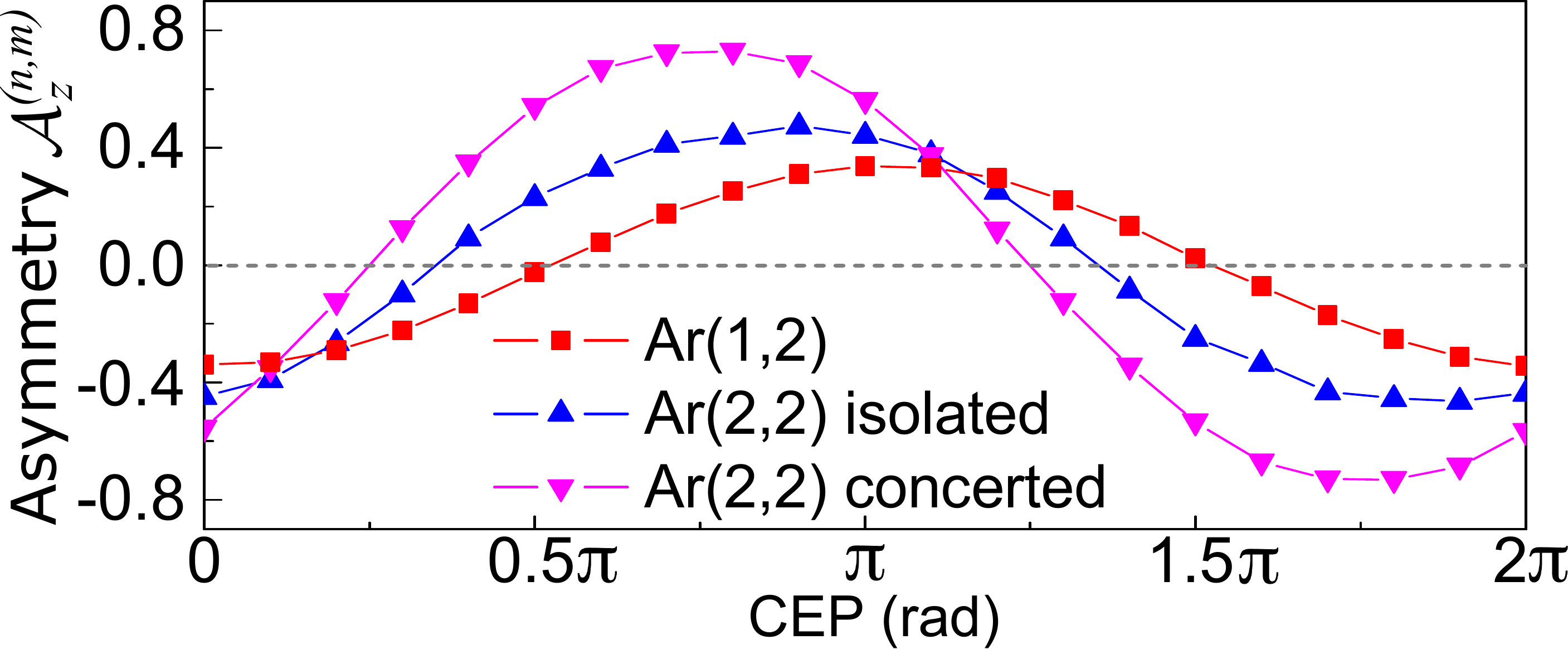}
\caption{Simulated \gls{cep}-dependence of the asymmetry $\asymm$ for channels Ar(1,2) and Ar(2,2).
The latter is separated into isolated and concerted two-electron emissions based on $\Delta t = t_{1st}-t_{A,B,C}$, where $t_{A,B,C}$ (indicated in Fig.~\ref{fig2}) are the instants of the laser field-maxima right before a given $t_{1st}$ peak that initiates the electron emission at $t_{1st}$: Isolated for $0\leq \Delta t \leq 0.25T$, concerted for $0.25T \leq \Delta t \leq 0.5T$.
\label{fig4}}
\end{figure}

\gls{lite} also plays a significant role in the concerted type of DI. Two cases can be distinguished: Representative electron trajectories [Figs.~\ref{fig3}(d,e)] show that for \textit{case 1} one of the electrons is emitted at one site and  transferred to the other site by \gls{lite}. There it is captured by the Coulomb potential, collides with the second electron initially on this site, and produces a doubly excited neutral atom, i.e., an Ar$^{2+}$-Ar$^{**}$ dimer. The highly excited Ar$^{**}$ atom is then doubly ionized before the next peak of the laser field, resulting in an  Ar(2,2) dimer.
\textit{Case 2} [Figs.~\ref{fig3}(f,g)] starts similarly: An electron is emitted at one site and is transferred to the other by \gls{lite}. However, in this case the energy exchange by collisions with the second electron is larger, so that one of the electrons gains enough energy to ionize soon. The other electron loses some of its energy and is trapped by the Coulomb potential, forming a transient Ar$^{2+}$-Ar$^{+*}$ complex. The captured electron finally ionizes at or after the next peak of the laser field and produces an Ar(2,2) dimer.

This second case is reminiscent of the recollision-induced excitation with subsequent field ionization (RESI) process well-known for monomers \cite{Feuerstein2001, Rudenko2004, Wang2016a}. Here, however, the collision-excitation step takes place on a distant entity and is enabled only by a preceding LITE process.
Further explanations and a  visualization of the role of LITE in the three different DI scenarios, as well as additional data and discussion on the role of the alignment of the argon dimer with respect to the laser polarization direction, the correlation between the two emitted electrons due to the collisions induced by LITE, and a spatio-temporal analysis of the electron transfer is provided in the \SM{}.


The finding that the DI dynamics to Ar(2,2) is dominated by an electron transfer process (LITE),  
explains why the second electron emission is delayed by a laser half-cycle to its initiating laser field-peak [cf. the $t_{2nd}$ distributions in Figs.~\ref{fig2}(f)-(h)]. 
Likewise, also the double-peak structure of $t_{1st}$ can be explained by LITE: In the concerted cases of DI the first electron is transferred and therefore is emitted with delay, giving rise to the second peak. The first, undelayed peak arises during the isolated cases of DI and during single ionization (SI) to Ar(1,2). The delayed peak in SI corresponds to cases where the first electron becomes transferred but the second electron stays bound, see \SM{} for further details. 

Finally, based on the fact that the first ionization step proceeds similarly for the Ar(1,2) and Ar(2,2) channels [cf. Figs.~\ref{fig2}(c)-(h)], we can now investigate which of the two DI cases, the isolated or the concerted one, is responsible for the distinct
\gls{cep}-shift observed between the $\asymm^{(1,2)}$ and $\asymm^{(2,2)}$ curves in Figs.~\ref{fig2}(a,b).
To see this, we plot in Fig.~\ref{fig4}  $\asymm^{(2,2)}$  separately for the isolated and concerted contributions to Ar(2,2), in comparison with $\asymm^{(1,2)}$ taken from Fig.~\ref{fig2}(b).
The separated curves reveal that the uncorrelated two-electron emission of the isolated case introduces a notable shift, but the main shift is introduced by the concerted pathway. The reason is that for this case the electron-electron interaction dynamics triggered in the excited argon atom upon electron-transfer by \gls{lite}  leads to electron emission over a much broader range of time within the laser cycle as compared to a purely field-driven ionization dynamics confined to around the crests of the laser cycle.

In conclusion, we have experimentally and theoretically studied the ionization-fragmentation dynamics of argon dimers in intense few-cycle laser pulses with a known \gls{cep}.
We observe a distinct CEP-shift of the electron emission asymmetry  between the $\text{Ar}^{+}+\text{Ar}^{2+}$ and $\text{Ar}^{2+}+\text{Ar}^{2+}$ fragmentation channels.
Using a classical ensemble model we find that this CEP-shift is due to electron-electron interaction mediated by a field-driven electron transfer process (\gls{lite}) from one argon atom to the other.
Our work, thus, heralds the possibility to use strong laser fields for controlling sub-cycle inter-molecular electron-transfer processes where the transferred electron can excite electronic dynamics on a distant molecule.  This finding opens up a new route for controlling molecular processes with intense laser pulses beyond mere bond-breaking reactions.



\acknowledgments
We thank Prof. Difa Ye and Jing Chen, and Dr. Zongqiang Yuan for stimulating discussion.
This work was supported by the Austrian Science Fund (FWF), Grants No.
P28475-N27,  
and P30465-N27,   
the National Key Research and Development Program of China (No. 2019YFA0307702), 
the National Natural Science Foundation of China (Nos. 11834015, 11847243, 11804374, 11922413 and 11874392), and the Strategic Priority Research Program of the Chinese Academy of Sciences (No. XDB21010400).


%

\end{document}